\documentclass[12pt]{article}
\pdfoutput=1
\usepackage{jheppub}

\usepackage[dvipsnames]{xcolor}

\usepackage{subcaption,caption}
\usepackage{amsmath}
\usepackage{amssymb}
\usepackage{amsfonts}
\usepackage{amsbsy}
\usepackage{bm}
\usepackage{color}
\usepackage{array}
\usepackage{dcolumn}
\usepackage{tensor}
\usepackage{braket}
\usepackage{eufrak}
\usepackage{mathrsfs}
\usepackage{slashed}
\usepackage{bbm}
\usepackage{bbold}
\usepackage{slashed}
\usepackage{tikz}
\usepackage{verbatim}

\newcommand{\ud}{\mathrm{d}}
\newcommand{\LCm}{{\scriptscriptstyle -}} 
\newcommand{\LCp}{{\scriptscriptstyle +}}

\newcommand{\LCperp}{{\scriptscriptstyle \perp}}

\newcommand{\mcA}{\mathcal{A}}

\newcommand{\mcD}{\mathcal{D}}

\newcommand{\mcM}{\mathcal{M}}

\newcommand{\mcO}{\mathcal{O}}

\newcommand{\mcS}{\mathcal{S}}

\newcommand{\mcV}{\mathcal{V}}

\newcommand{\mcX}{\mathcal{X}}
\newcommand{\mcY}{\mathcal{Y}}

\begin{document}

\title{
The analytic structure of amplitudes on backgrounds from gauge invariance and the infra-red
}

\author{Anton Ilderton}
\emailAdd{anton.ilderton@plymouth.ac.uk}

\author{and Alexander J.~MacLeod}
\emailAdd{alexander.macleod@plymouth.ac.uk}

\affiliation{Centre for Mathematical Sciences, University of Plymouth, PL4 8AA, UK}

\abstract{
Gauge invariance and soft limits can be enough to determine the analytic structure of scattering amplitudes in certain theories. This prompts the question of how gauge invariance is connected to analytic structure in more general theories. Here we focus on QED in background plane waves. We show that imposing gauge invariance introduces new virtuality poles into internal momenta on which amplitudes factorise into a series of terms. Each term is gauge invariant, has a different analytic structure in external momenta, and exhibits a hard/soft factorisation. The introduced poles are dictated by infra-red behaviour, which allows us to extend our results to scalar Yukawa theory. The background is treated non-perturbatively throughout. 
}

\maketitle

\section{Introduction}
%
It has been shown that gauge invariance is enough to completely determine scattering amplitudes and their underlying analytical structure in certain theories~\cite{Barreiro:2013dpa,Arkani-Hamed:2016rak,Boels:2016xhc,Berg:2016fui,Boels:2017gyc,Fu:2017uzt,Barreiro:2019ncv}, and it has been conjectured that locality and unitarity emerge as a consequence of imposing gauge invariance~\cite{Arkani-Hamed:2016rak, Rodina-Adler}. The investigation of which principles determine scattering amplitudes is not limited to gauge theories; it has been shown that soft theorems are enough to fix tree-level scattering amplitudes in the non-linear sigma model and Dirac-Born-Infeld~\cite{Cheung:2015ota,Rodina:2018pcb}, and to impose strong constraints on the Lagrangians of both scalar and vector effective field theories~\cite{Cheung:2014dqa,Padilla:2016mno,Cheung:2018oki}.

While the majority of theories considered in this context share the property of being massless, similar results in very different theories point to an underlying structure or principle~\cite{Cheung:2017ems,Carrasco:2019qwr}, and one can ask to what extent gauge invariance and soft theorems fix behaviour in theories with coupling to matter~\cite{Bonifacio:2019rpv} or in other sectors of the standard model~\cite{Durieux:2019eor,Bachu:2019ehv}. The question we investigate here is to what extent gauge invariance and soft/infra-red behaviour can be exploited to uncover the underlying analytic structure of amplitudes in background fields.

Given that an arbitrary background (coupling to some set of fields in a theory) introduces an arbitrary amount of additional structure, it is not obvious if/how gauge invariance could (fully) determine properties of amplitudes in that background. We will find, though, that traces of the above results on gauge invariance and soft limits do persist.  We consider QED with an additional background electromagnetic field. We will show, using tree-level amplitudes in the background, that imposing explicit gauge invariance uncovers a hidden analytic structure; gauge invariance demands a certain infra-red behaviour which introduces new poles in the  internal momenta. These poles affect the analytic structure of the \textit{entire} amplitude (not just the infra-red part); the amplitude factorises on the internal poles with the residues being \textit{individually gauge-invariant sub-amplitudes},  each with distinct  analytic structures in the \textit{external}, scattered, momenta.

The connection between gauge invariance of amplitudes and the infra-red allows us to extend our results to theories without gauge invariance. We will show for a simple scalar Yukawa theory that the infra-red structure of amplitudes leads to an almost identical factorisation of scattering amplitudes. 

Our chosen background is an electromagnetic (or later scalar) ``sandwich'' plane wave of finite extent. Here, the high degree of symmetry frequently allows exact solutions~\cite{volkov35,Levy1965,Duval:2017els,Adamo:2017sze}, and our results will be exact in the coupling to the background. The same background has been used to test the ``double copy'' conjecture (for a review see \cite{Bern:2019prr}) beyond flat spacetimes~\cite{Adamo:2017nia,Adamo:2018mpq}.

An outline of our results is as follows. Consider a tree-level four-point QED amplitude in an external field, where all external particles are fermions and hence there is an internal photon line.  The corresponding amplitude is defined in position space, due to a nontrivial dependence of the background on position. For the case of plane waves, there is at each vertex a nontrivial dependence on a single spacetime coordinate $x^\LCp := n\cdot x$ for some lightlike vector $n^\mu$. As such only \textit{three}  momentum components are conserved at each vertex, and overall. Stripping off the $\delta$-function conserving overall three-momentum, the amplitudes $\mcM$ for our processes may be written in the form
\begin{equation}\label{eqn:GeneralAmplitude}
\mcM
\sim
\int
\!
\ud v
\;
\mcA^\nu_\mcY(v)
\frac{\widetilde{D}_{\mu\nu}}{v + i \epsilon}
\mcA^\mu_\mcX(v)
\;, 
\end{equation}
in which $\widetilde{D}$ is the tensor structure of the photon propagator in some gauge, $v$ is the photon virtuality, and the amplitude naturally factorises at the on-shell pole $v=0$ into two sub-amplitudes, call them $\mcA_\mcX$ and $\mcA_\mcY$. These are given by nontrivial spacetime integrals over $x^{\LCp}$ dependence at three-point vertices, which are not analytically computable in general. The sub-amplitudes both have a structure
\begin{equation}\label{eqn:GeneralVertexGauge}
\mcA^\mu_{i}(v)
\sim
\int
\!
\ud x^\LCp
\,
\big[
\mcV_{0}^\mu + \mcV^\mu(x^\LCp)
\big]
e^{
	i 
	\Phi(x^\LCp;v)
}
\;,
\end{equation}
in which $\mcV_0$, $\mcV(x^\LCp)$ and $\Phi(x^\LCp;v)$ take different forms at each vertex, but their important properties are common; $\mcV(x^\LCp)$ depends on the background while $\mcV_{0}$ does not and so $\mcV_{0}$ multiplies a pure phase term depending on $\Phi(x^\LCp;v)$, which is linear in $v$.  It is then clear that the virtuality integral in (\ref{eqn:GeneralAmplitude}) could be performed \textit{before} the spacetime integrals at the vertices. This  is what is normally done in the literature on QED scattering in intense fields modelled as plane waves (for connections to which see Appendix \ref{app:Trident}); one either separates the virtuality factor into a $\delta$-function and principal value (both of which contribute since the internal line can go on-shell in a background) or performs the $v$-integral directly via contour integration~\cite{Ritus:1972nf,ilderton11,seipt12,King:2013osa}. The two methods lead to different representations of the amplitude with different physical interpretations. A similar issue arises with the choice of gauge for $\widetilde{D}_{\mu\nu}(\ell)$ in (\ref{eqn:GeneralAmplitude}); each choice yields a different division of terms, requiring results to be cross-checked to ensure gauge invariance is preserved~\cite{dinu18,mackenroth18}.

We do something different. The key observation is that the amplitude (\ref{eqn:GeneralAmplitude}) is not, as we will see, \textit{manifestly} gauge invariant. It is known how to resolve this in the approaches cited above, but in contrast we address the issue before proceeding with the calculation. We will show that if gauge invariance is imposed first then additional poles are introduced into the sub-amplitudes, so (\ref{eqn:GeneralVertexGauge}) becomes
\begin{equation}\label{eqn:Bob}
\mcA^\mu_{i}(v)
~
\longrightarrow
~
\int
\!
\ud x^\LCp
\Big[
\sum_j
\frac{\Delta_j}{v - v_j \pm i \epsilon}
\mcV_{0}^\mu 
+ 
\mcV^\mu(x^\LCp)
\Big]
e^{
	i 
	\Phi_{i}(x^\LCp;v)
}
\;,
\end{equation}
in which the pure phase term has acquired a series of new poles $v_j$ in the virtuality $v$, and additional factors $\Delta_j$ in the corresponding residues. This new structure renders the sub-amplitudes individually gauge invariant. Upon performing the virtuality integral in (\ref{eqn:GeneralAmplitude}), the full amplitude now factorises not just on the usual $v=0$ pole but also on (combinations of) each of the internal poles. Remarkably, we will find that each term in this factorisation is individually gauge invariant and has a different analytic structure in the \textit{external}  momenta. In deriving these results we will see that ensuring gauge invariance is intimately connected to the infra-red, or large distance, behaviour of the phase terms appearing in (\ref{eqn:GeneralVertexGauge}), the poles, and the pole prescriptions in (\ref{eqn:Bob}).  As a result, our new representation of the amplitude (\ref{eqn:GeneralAmplitude}) will exhibit a factorisation of soft terms. It is this connection to the infra-red which will also allow us to uncover similar structures in non-gauge theories.

This paper is organised as follows. In Sect.~\ref{sec:QED} we first introduce QED scattering calculations in background plane waves. We explain how gauge invariance of amplitudes leads to the appearance of new poles in internal momenta. We then evaluate the amplitude in this form and highlight its important structures, in particular its dependence on external momenta.  In Sect.~\ref{SECT:DISCUSS} we investigate the decomposition of our amplitude in detail, identifying in them a background-field dependent generalisation of soft/hard factorisation.  In Sect.~\ref{sec:Scalar} we extend our results to a simple scalar Yukawa interaction, where the infra-red behaviour leads to an analogous decomposition and factorisation. We conclude in Sect.~\ref{sec:outro}.

%
%
\section{QED amplitudes: gauge invariance and the infra-red\label{sec:QED}}
%
%
\subsection{Scattering on plane wave backgrounds}
We work in lightfront coordinates $x^\mu=(x^\LCp,x^\LCm,x^\LCperp)$ with $\ud s^2 =\ud x^\LCp \ud x^\LCm - \ud x^\LCperp \ud x^\LCperp$ and $\perp =1,2$. (Our results extend directly to $d>4$ dimensions.)  These coordinates match the symmetry properties~\cite{Levy1965,Duval:2017els,Zhang:2019gdm} of our plane wave background, defined by
\begin{equation}\label{Rosen}
eA  = a_\LCperp (x^\LCp) \ud x^\LCperp \;.
\end{equation}
The electromagnetic fields of the background are $E^\LCperp = -a'_\LCperp$ and $B^\LCperp = \epsilon^{\LCperp j}a_j'~(j=1,2)$. We  consider `sandwich' plane waves for which the electromagnetic fields vanish as $x^\LCp\to \pm\infty$; this splits spacetime into causally separated flat and non-flat regions~\cite{Bondi:1958aj} and gives good scattering boundary conditions in `lightfront time' $x^\LCp$. We can always fix $a_\LCperp(-\infty)=0$. Using the `Einstein-Rosen'~\cite{Monteiro:2014cda,Adamo:2017nia} gauge (\ref{Rosen}) makes the physics manifest, as the classical momentum of an electron, charge $e$, entering the wave from $x^\LCp = -\infty$ with momentum $p_\mu$ may be expressed directly in terms of $a_\mu \equiv \delta_{\mu}^{\LCperp} a_{\LCperp}$ as
\begin{equation}\label{pi-1}
\pi_\mu(x^\LCp) = p_\mu - a_\mu(x^\LCp) + \frac{2 p \cdot a(x^\LCp) - a^2(x^\LCp)}{2 n \cdot p}
n_\mu \;,
\end{equation}
in which $n_\mu$ is defined by $n \cdot x = x^\LCp$. We write $\hat{\pi} := \pi(-a)$ for positrons. Note that $\pi^2=p^2=m^2$, on-shell. It is clear from (\ref{pi-1}) that particle propagation in plane waves can exhibit a memory effect~\cite{Ehlers:1962zz,Dinu:2012tj,Zhang:2017rno,Hamada:2018cjj,Shore:2018kmt} if $a_\LCperp(\infty)$ is nonvanishing~\cite{Dinu:2012tj}. For the sake of simplicity we set $a_\LCperp(\infty)=0$ here; only minor extensions, amounting to~slightly modified LSZ rules~\cite{Kibble:1965zza,Dinu:2012tj}, are needed to extend our results to the general case. 

Amplitudes in plane waves are calculated using background perturbation theory~\cite{DeWitt:1967ub,tHooft:1975uxh,Boulware:1980av,Abbott:1981ke,Furry51}: the background is treated exactly, while scattering of (matter and) photons is treated as a perturbation around the background. Practically this means, in the path integral, expanding in the coupling $e$ as usual while treating $a_\mu$ exactly (non-perturbatively) as part of the `free' action. Such calculations can be performed explicitly in plane waves due to their many symmetries~\cite{volkov35,Levy1965,Duval:2017els}. The position space Feynman rules are as follows. The vertex is $-ie \gamma^\mu$ as usual and the photon propagator is
\begin{equation}\label{def:PhotonProp}
-iD_{\mu\nu}(x - y)
=
-i\int
\!
\frac{\ud^4 \ell}{(2\pi)^4}
\frac{\widetilde{D}_{\mu\nu}}{\ell^2 + i \epsilon}
\,
e^{-i\ell\cdot (x-y)} \;,
\end{equation}
in which we leave $\widetilde{D}_{\mu\nu}$ unspecified so that we may work in an arbitrary gauge. Incoming/outgoing photons of momentum $\ell_\mu$ and polarisation $\varepsilon_\mu$ are described by $\varepsilon_\mu e^{\mp i (\ell \cdot x)}$ where $\varepsilon \cdot \ell=0$ as usual.  The fermion propagator $S_V(x,y)$ is now `dressed', being given by the inverse of the background covariant derivative $i\slashed{\partial} - \slashed{a} - m $:
\begin{equation}\label{VolkProp}
S_V(x,y)
=
\int\!
\frac{\ud^4 q}{(2\pi)^4}
\bigg(
1
+
\frac{\slashed{a}(y^\LCp)\slashed{n}}{2 n \cdot q}
\bigg) 
\frac{\slashed{q} + m}{q^2 - m^2 + i \epsilon}
\bigg(
1
+
\frac{\slashed{n} \slashed{a}(x^\LCp)}{2 n \cdot q}
\bigg) 
e^{- i S_q(x) + i S_q(y)} \;,
\end{equation}
in which $S_p$ is the classical action of a particle in the plane wave,
\begin{equation}
S_p(x)
\equiv
p \cdot x
+
\int\limits^{x^\LCp}_{-\infty}
\frac{2 p \cdot a - a^2}{2 n \cdot p} \;.
\end{equation}
LSZ reduction of the propagator (\ref{VolkProp}) yields the ``Volkov wavefunctions'' for external fermion legs~\cite{volkov35}. These describe initially free fermions propagating from the `in' region of spacetime (causally before the sandwich plane wave switches on) to the `out' region (after it has switched off)~\cite{Bondi:1958aj,Heinzl:2000ht}. For incoming electrons the Volkov wavefunction~is
\begin{equation}\label{def:Volkov}
\Psi_p(x)
=
\bigg(
1
+
\frac{\slashed{n} \slashed{a}(x^\LCp)}{2 n \cdot p}
\bigg) 
u_p e^{-i S_p(x)} = u_\pi(x^\LCp)
e^{-i S_p(x)}
\;,
\end{equation}
where $u_\pi$ is just a standard $u$-spinor for the on-shell momentum $\pi_\mu$ in~(\ref{pi-1}).  The scalar part of $\Psi_p$ reproduces the momentum $\pi_\mu$ when acted on with the background-covariant derivative:
\begin{equation}
i
\mcD_\mu
e^{- i S_p(x)}
= \pi_\mu(x^\LCp) e^{-i S_p(x)} \;.
\end{equation}
Outgoing electrons are described by $\bar{\Psi}_{p}$ with $-\infty \to \infty$ in the integral limit, and incoming/outgoing positrons similarly by $\bar{\Psi}_{-q}$/$\Psi_{-q}$. In the limit of vanishing background $a_\mu(x^\LCp) \rightarrow 0$, $\Psi_p$ reduces to the usual free particle wavefunction $u_p e^{-ip.x}$. Observe that (\ref{VolkProp}) and (\ref{def:Volkov}) are exact for \textit{any} value of the dimensionless effective coupling to the background $\sim a/m$, even $a/m\gg 1$; for applications see~\cite{DiPiazza2012a,King:2015tba,Seipt:2017ckc}.

\subsection{$4$-point amplitudes}
\begin{figure}
	\centering
	\includegraphics[width=0.2\textwidth,trim={5.1cm 19cm 13cm 4cm},clip=true]{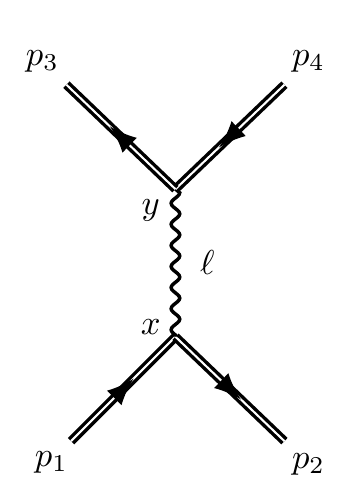} \qquad\qquad\qquad
	\includegraphics[width=0.2\textwidth,trim={5.1cm 19cm 13cm 4cm},clip=true]{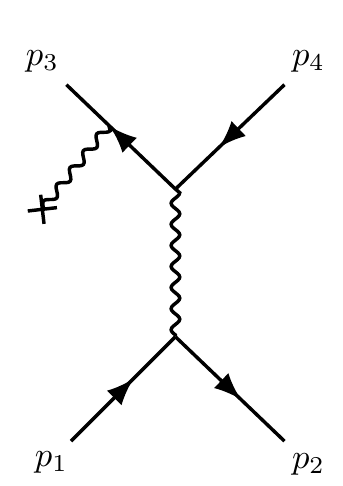}
	\caption{\label{fig:Trident} \textit{Left}: the tree level $e^- e^+ \to e^- e^+$ amplitude (\ref{eqn:TridentAmplitude}) in a plane wave, where double lines represent the wavefunctions (\ref{def:Volkov}) which include \textit{all orders} of interaction with the background. \textit{Right:} one of the (four) lowest order, five-point contributions to the same process, calculated perturbatively in the background, indicated by a photon line connected to cross.
	}
\end{figure}
We consider four-point fermion amplitudes as shown in Fig.~\ref{fig:Trident}, which is already enough to demonstrate our results. In particular consider electron-positron scattering,
\begin{equation}\label{eqn:Trident}
e^-(p_1)
+
e^+(p_2)
\rightarrow 
e^{-}(p_3)
+
e^{+}(p_4)
,
\end{equation}
where $p_j^2 = m^2$. The tree level scattering amplitude $\mcS$ for this process is, in terms of the Volkov functions (\ref{def:Volkov}) and the photon propagator~$D_{\mu\nu}$,
\begin{equation}\label{eqn:TridentAmplitude}
\mcS
=
i
e^2
\int
\! 
\ud^4 x
\,
\ud^4 y
\;
\bar{\Psi}_{p_3}(y)
\gamma^\mu
\Psi_{-p_4}(y)
\,
D_{\mu\nu}(y - x)
\,
\bar{\Psi}_{-p_2}(x)
\gamma^\nu
\Psi_{p_1}(x)
\,\,\,  + \,\,\, \ldots\;.
\end{equation}
The ellipses represent the other interaction channels -- for brevity we consider only the $s$-channel diagram in Fig.~\ref{fig:Trident}, but all our discussions apply equally to $t$ and $u$ channels and to other processes by swapping external legs. At any vertex in a plane wave background the integrals over $\{x^-, x^\LCperp\}$ can be carried out as usual to yield conservation of the \textit{three} momentum components $p^\LCp$ and $p^\LCperp$. As such $\mcS$ has the form
\begin{equation} \label{def:Smatrix}
\mcS
=
e^2
(2\pi)^3 \delta^3_{LF}(p_4 + p_3 - p_2 - p_1)
\mcM \;,
\end{equation}
where $\delta^3_{LF}(p) \equiv \delta(p^\LCp)\delta^2(p^\LCperp)$. Three components of the internal photon momenta $\ell_\mu$ are fixed by momentum conservation, so from here $\ell^\mu = \ell^\mu_\star + v n^\mu$ in which
\begin{equation}\label{def:lstar}
\ell^\mu_\star 
= 
p_1^\mu + p_2^\mu - \frac{(p_1 + p_2)^{2}}{2 n \cdot (p_1 + p_2)} n^\mu
\,\, =
\,\, p_3^\mu + p_4^\mu - \frac{(p_3 + p_4)^2}{2 n \cdot (p_3 + p_4)} n^\mu \;,
\end{equation}
is on-shell ($\ell_\star^2 = 0$) and $v$ is the photon virtuality. Thus the reduced amplitude $\mcM$ contains an integral over the virtuality $v$ and nontrivial integrals over $x^\LCp$ and $y^\LCp$ due to the spacetime dependence of the Volkov wavefunctions. It takes the form
\begin{equation}\label{eqn:AmplitudeGaugeDependent}
\mcM
=
\frac{i}{2 n \cdot \ell_\star}
\!
\int
\!
\frac{\ud v}{2\pi}
\;
\mcA^\mu_{\mcY}(v)
\frac{\widetilde{D}_{\mu\nu}}{v + i \epsilon}
\mcA^\nu_{\mcX}(v) \;,
\end{equation}
in which the two sub-amplitudes for pair annihilation and pair creation at the spacetime points $x$ and $y$ respectively are, 
\begin{equation}\label{eqn:ComptonVertexGauge}
\mcA^\mu_{\mcX}(v)
=
\int
\!
\ud x^\LCp
\Big[
\mcX_0^\mu + \mcX^\mu(x^\LCp)
\Big]
e^{
	i 
	\Phi_{\mcX}(x^\LCp;v)
}
\;, \qquad
\mcA^\mu_{\mcY}(v)
=
\int
\!
\ud y^\LCp
\Big[
\mcY_0^\mu + \mcY^\mu(y^\LCp)
\Big]
e^{
	i
	\Phi_{\mcY}(y^\LCp;v)
}
\;,
\end{equation}
with $\mcX_0^\mu = \bar{v}_{p_2} \gamma^\mu u_{p_1}$ and $\mcY_0^\mu
= \bar{u}_{p_3} \gamma^\mu v_{p_4}$ the background-free spin structures at the vertices, and  $\mcX^\mu(x^\LCp)$ and $\mcY^\mu(y^\LCp)$ the background-dependent parts,  
\begin{align}
\label{def:ComptonSpinField}
&
\mcX^\mu(x^\LCp)
= 
\frac12
\bar{v}_{p_2}
\bigg[
\frac{\gamma^\mu\slashed{n} \slashed{a}}{n \cdot p_1}
-
\frac{\slashed{a}\slashed{n}\gamma^\mu}{n \cdot p_2}
+
\frac{
	a^2 n^\mu \slashed{n}
}{n \cdot p_1 \, n \cdot p_2}
\bigg]
u_{p_1}
\;, \\
&
\mcY^\mu(y^\LCp)
=
\frac12
\bar{u}_{p_3}
\bigg[
\frac{
	\slashed{a}
	\slashed{n}
	\gamma^\mu
}{n \cdot p_3}
-
\frac{
	\gamma^\mu
	\slashed{n} 
	\slashed{a}
}{n \cdot p_4}
+
\frac{
	a^2
	n^\mu
	\slashed{n} 
}{n \cdot p_3\, n \cdot p_4}
\bigg]
v_{p_4}
\;,
\end{align}
(suppressing for conciseness the dependence of the background on $x^\LCp$ or $y^\LCp$)  and the phase functions in the exponents are, writing $\pi_1 := \pi(p_1)$ etc, 
\begin{equation}\label{def:ComptonPhase}
\Phi_{\mcX}(x^\LCp;v)
=
\int\limits^{x^\LCp}
v
-
\frac{\ell_\star \cdot (\pi_1 + \hat{\pi}_2)}{n \cdot (p_1 + p_2)}
\;, \quad
\Phi_{\mcY}(y^\LCp;v)
=
\int\limits^{y^\LCp}
\frac{\ell_\star \cdot (\pi_3 + \hat{\pi}_4)}{n \cdot (p_3 + p_4)}
-
v
\;.
\end{equation}
Despite the complexity, the essential properties of these objects are simply that $\mcX^\mu_0$ and $\mcY^\mu_0$ are constants, $\mcX^\mu(x^\LCp)$ and $\mcY(y^\LCp)$ vanish outside the sandwich wave, and the phase functions $\Phi$ are linear in $x^\LCp$/$y^\LCp$ both causally before and after the sandwich wave.

\subsection{Gauge invariance and the infra-red}
%
The $4$-point amplitude (\ref{eqn:AmplitudeGaugeDependent}) is not explicitly gauge invariant\footnote{This is not due to neglecting other channels  -- the individual diagrams should be invariant here.}. To see this, make the replacement $\widetilde{D}_{\mu\nu} \rightarrow \ell_\mu q_\nu(\ell) + \ell_\nu q_\mu(\ell)$, for $q_\mu(\ell)$ an arbitrary function of~$\ell_\mu$; the amplitude $\mathcal A$ should then vanish, but does not. We expect that $\ell_\mu$ dotted into one of the sub-amplitudes should vanish, so $\ell \cdot \mcA_{\mcX}(v) = \ell \cdot \mcA_{\mcY}(v) = 0$, but instead one finds
\begin{equation*}
\ell \cdot \mcA_\mcX(v) 
=
-i \bar{v}_{p_2}
\slashed{n} 
u_{p_1} 
\,
\int
\!
\ud x^\LCp \frac{\ud}{\ud x^\LCp}
e^{
	i 
	\Phi_{\mcX}(x^\LCp;v)
}
\;, \quad
\ell \cdot \mcA_{\mcY}(v)
=
i\bar{u}_{p_3}
\slashed{n}
v_{p_4}
\int
\!
\ud y^\LCp
\frac{\ud}{\ud y^\LCp}
e^{
	i
	\Phi_{\mcY}(y^\LCp;v)
}
.
\end{equation*}
These are boundary terms~\cite{UsAbsorption}, but they are ambiguous since the pure phases oscillate without damping asymptotically.  Gauge invariance is thus closely tied to the \textit{infra-red} behaviour of the sub-amplitudes, and we must make the latter explicit in order to ensure that the former is preserved -- it is here that our calculation deviates from the usual route taken in the literature. To expose the infra-red behaviour and its consequences, we take the phase integral and insert as usual convergence factors $\exp(-\epsilon|x^\LCp|)$~\cite{Weinberg:1995mt,Peskin:1995ev} -- we can w.l.o.g. take the sandwich wave to switch on at $x^\LCp=0$ and off at $x^\LCp=T>0$. Using the pure phase term in $\mcA_{\mcX}$ to illustrate, the integral to consider is,
\begin{equation}
\int
\!
\ud x^\LCp
\,
e^{
	i 
	\Phi_{\mcX}
}
\to 
\int_{-\infty}^{0}
\!
\ud x^\LCp
\,
e^{
	i 
	\Phi_{\mcX}
	+
	\epsilon x^\LCp
}
+
\int_{0}^{T}
\!
\ud x^\LCp
\,
e^{
	i 
	\Phi_{\mcX}
	-
	\epsilon x^\LCp
}
+
\int_{T}^{+\infty}
\!
\ud x^\LCp
\,
e^{
	i 
	\Phi_{\mcX}
	-
	\epsilon x^\LCp
}
\;.
\label{217}
\end{equation}
The outer integrals can be performed exactly, as $\Phi_\mcX$ is linear in
$x^\LCp$ outside of the background. For the inner integral we integrate by parts
once to generate terms which cancel the boundary terms from the outer integrals,
and then integrate by parts again, using that $a(0)=a(T)=0$, to put (\ref{217})
in the form
\begin{equation}\label{eqn:CutIntegral}
\int
\!
\ud x^\LCp
\,
e^{
	i 
	\Phi_{\mcX}
}
=
i
\bigg[
\frac{1}{v - v_\star + i \epsilon}
-
\frac{1}{v - v_\star - i \epsilon}
\bigg]
-
\frac{v_\star}{v - v_\star + i \epsilon}
\int
\!
\ud x^\LCp
\,
\Delta_\mcX(x^\LCp)
\,
e^{
	i 
	\Phi_\mcX
} \;,
\end{equation}
where we have defined 
\begin{equation}\label{def:DeltaC}
v_\star
=
\frac{(p_1 + p_2)^2 }{2 n \cdot (p_1 + p_2)}
\;,
\qquad \Delta_\mcX(x^\LCp)
=
1  
-
\frac{\ell_\star \cdot (\pi_1(x^\LCp) + \hat{\pi}_2(x^\LCp))}{\ell_\star \cdot (p_1 + p_2) }
\;.
\end{equation}
Gauge invariance has therefore given us, via a standard infra-red regularisation~\cite{Weinberg:1995mt,Peskin:1995ev}, a better-defined expression for the pure phase integral. Writing the sum of poles in the square brackets as $2\pi\delta(v - v_\star)$ we see that this term is just the background-free result, while  the integrand of the second term in (\ref{eqn:CutIntegral}) vanishes outside the sandwich  wave because the scalar factor $\Delta_\mcX(x^\LCp)$ goes to zero for $a\to 0$.  The essential point is that the same phase integral as in (\ref{eqn:CutIntegral}) appears in the sub-amplitude $\mcA_\mcX$; thus we have 
\begin{equation}
\label{eqn:RegularisedCompton}
\mcA^\mu_{\mcX}(v)  
\to
2 \pi
\delta(v - v_\star)\,
\mcX_0^\mu 
+
\int \!\ud x^\LCp
\, e^{i \Phi_\mcX(x^\LCp;v)} 
\bigg[
\frac{- v_\star}{v - v_\star + i \epsilon}
\Delta_\mcX(x^\LCp)\mcX_0^\mu 
+ \mcX^\mu(x^\LCp)\bigg]
\;,
\end{equation}
With this regulated expression for $\mcA_\mcX$ we can verify directly that $\ell \cdot \mcA_\mcX = 0$, with no ambiguous boundary term. Repeating the calculation for the pair production vertex, gauge invariance of the \textit{full} amplitude $\mathcal{M}$ becomes manifest. We then have
\begin{align}\label{eqn:RegularisedAmplitude}
&
\mcM
=
\frac{i}{2 n \cdot \ell_\star}
\int
\!
\frac{\ud v}{2\pi} \, \frac{1}{v + i \epsilon}
\widetilde{D}_{\mu\nu}
\nonumber\\
&
\bigg(
2\pi \delta(v - \bar{v}_\star)
\mcY_0^\mu
+
\!
\int\!
\ud y^\LCp \, 
e^{
	i
	\Phi_\mcY(y^\LCp;v)
}
\bigg[
\frac{-\bar{v}_\star}{v - \bar{v}_\star - i \epsilon} 
\Delta_\mcY(y^\LCp)
\mcY_0^\mu
+ 
\mcY^\mu(y^\LCp)
\bigg]
\bigg) 
\nonumber\\
&\bigg(
2 \pi
\delta(v - v_\star)
\mcX_0^\nu 
+
\!\!
\int\!
\ud x^\LCp \, 
e^{i \Phi_\mcX(x^\LCp;v)} 
\bigg[
\frac{- v_\star}{v - v_\star + i \epsilon}
\Delta_\mcX(x^\LCp)\mcX_0^\nu 
+ \mcX^\nu(x^\LCp)\bigg]
\bigg)
\end{align}
in which the first line contains the gauge invariant pair production vertex with
\begin{equation}\label{def:tstar}
\qquad 
\bar{v}_\star
=
\frac{(p_3 + p_4)^2 }{2 n \cdot (p_3 + p_4)} \;,
\qquad
\Delta_\mcY(y^\LCp)
=
1
-
\frac{\ell_\star \cdot (\pi_3(y^\LCp) + \hat{\pi}_4(y^\LCp))}{\ell_\star \cdot (p_3 + p_4) }
\;.
\end{equation}
What we highlight is that imposing gauge invariance, through regularising the infra-red behaviour of the amplitude, uncovers additional poles in the virtuality at $v= v_\star$ and $\bar v_\star$, not present in (\ref{eqn:AmplitudeGaugeDependent})--(\ref{eqn:ComptonVertexGauge}) where there is only the propagator pole at $v = 0$. When we integrate over $v$, the poles will affect not just the infra-red part of amplitude, but the analytic structure of the \textit{whole} amplitude when considered as a function of external momenta.

\subsection{Gauge invariant factorisation at the poles}
%
Expanding out (\ref{eqn:RegularisedAmplitude}) yields several terms with different sets of virtuality poles.  Integrating over $v$ then picks up the residues from each set of poles, at which the whole amplitude factorises into a pair annihilation part and a pair production part.

\begin{figure}[t!]
	\centering\includegraphics[width=0.5\textwidth,trim={5.0cm 14.5cm 5.0cm 4cm},clip=true]{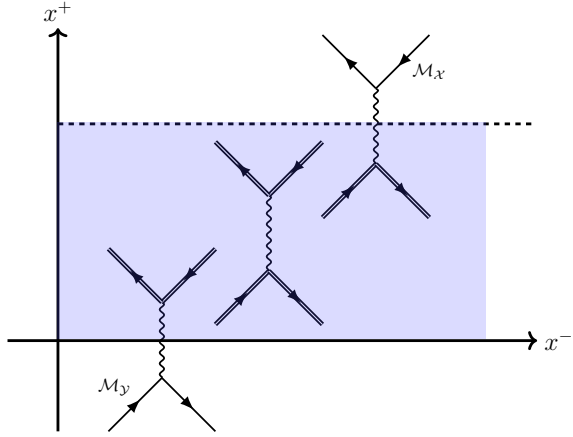}
	\caption{\label{fig:sketch} Illustration of some terms in the decomposition (\ref{eqn:AmplitudeDecomposition}). The shaded region indicates the sandwich plane wave field. One vertex in the terms $\mcM_\mcX$ and $\mcM_\mcY$ effectively lies outside the field, and so is represented by background-free vertices (single lines). The terms $\mcM^{\text{on}}$, $\mcM^\uparrow$ and $\mcM^\downarrow$, are dressed (double lines) at each vertex, however the way in which each vertex interacts with the background is distinct (see the text).}
\end{figure}

The sub-amplitudes $\mcA_\mcX$ and $\mcA_\mcY$ are themselves made up of terms with different numbers of poles, so integrating over $v$ will split them up; na\"{i}vely, this would appear to be a disadvantage given that their form is set by gauge invariance. However, we find that the pole structure is such that each resulting term is fully gauge-invariant and, furthermore, that each term also has a different analytic structure in the \textit{external} momenta.  There are six terms,
\begin{equation}\label{eqn:AmplitudeDecomposition}
\mcM
=:
\mcM^{\text{vac}}
+
\mcM^{\text{on}}
+
\mcM_{\mcX}
+
\mcM_{\mcY}
+
\mcM^\uparrow
+
\mcM^\downarrow
\;.
\end{equation}
which we consider in order. To simplify notation it is convenient to define the sum of two momenta $p_i$ and $p_j$ as
\begin{equation}
P_{ij} := p_i + p_j \;,
\end{equation}
in what follows. The first thing we learn about the decomposition (\ref{eqn:AmplitudeDecomposition}) is that it separates off the vacuum contribution to the total amplitude. $\mcM^\text{vac}$ comes from the product of $\delta$-functions in (\ref{eqn:RegularisedAmplitude}) and gives the usual S-matrix element for $e^- e^+ \rightarrow e^- e^+$ without background; reinstating the momentum $\delta$-function in (\ref{def:Smatrix}), we have
\begin{equation}\label{eqn:Vacuum}
\mcS^{\text{vac}}
=
ie^2
(2\pi)^4
\delta^4\big(P_{12} - P_{34}\big)
\,
\frac{\mcY_0 \cdot \mcX_0}{P_{12}^2}
\;.
\end{equation}
The second term $\mathcal{M}^\text{on}$ in (\ref{eqn:AmplitudeDecomposition}) picks up only the propagator pole at zero virtuality, $v=0$, which puts the internal line on-shell, $\ell \to \ell_\star$ introduced above. Explicitly,
\begin{align}\label{eqn:ComptonPairsOn}
\mcM^{\text{on}}
=
&
\frac{1}{2 n \cdot \ell_\star}
\int
\!\ud y^\LCp\!
\!\int^{y^\LCp}
\!
\!
\!
\!
\ud x^\LCp
\nonumber\\
&
\times
\;
e^{
	i
	\Phi_\mcY(y^\LCp;0)}
\big[
\Delta_\mcY(y^\LCp)
\mcY_0
+ 
\mcY(y^\LCp)
\big]
\cdot
\big[
\Delta_\mcX(x^\LCp)\mcX_0 
+ \mcX(x^\LCp)
\big] e^{i \Phi_\mcX(x^\LCp;0)} 
\;.
\end{align}
This term comprises two complete, regulated vertices (evaluated at  $v=0$), and is manifestly gauge invariant, hence we have replaced $\widetilde{D}_{\mu\nu} \to \eta_{\mu\nu}$. The time-ordering, which follows from the residue theorem, enforces causality for the real photon:  pair annihilation occurs before pair production. The integrals extend only over the sandwich wave duration (otherwise the integrand vanishes), so both pair annihilation and production occur within the field. This is illustrated in Fig.~\ref{fig:sketch}.

In all remaining terms of (\ref{eqn:AmplitudeDecomposition}) the intermediate photon is off-shell. The next term $\mcM_{\mcX}$ factorises at the poles at $v = \bar{v}_\star$ (which were combined into a $\delta$-function),
\begin{equation}\label{eqn:ComptonVacuum}
\mcM_{\mcX}
=
\frac{ 
	i
}{P_{34}^2}
\mcY_0
\cdot
\int
\!
\ud x^\LCp
\bigg[
\frac{P_{12}^2}{P_{12}^2- P_{34}^2 }
\Delta_\mcX(x^\LCp)\mcX_0
+ \mcX(x^\LCp)\bigg]
e^{i \Phi_\mcX(x^\LCp;\bar{v}_\star)} \;.
\end{equation}
There is now only a single integral; the regularised annihilation vertex lies within the field. The pair production vertex, though, has reduced to the vacuum vertex $\mcY_0^\mu$ defined below (\ref{eqn:ComptonVertexGauge}). Further, the pole sets the internal photon momentum to $\ell = P_{34}$ i.e.~this part of the amplitude obeys free-space conservation of \textit{four}-momentum at the pair production vertex (hence the leading factor of $1/P_{34}^2$). In other words, the pair production vertex effectively lies outside the field, see Fig.~\ref{fig:sketch}. Further, having picked up a different pole, the denominator of (\ref{eqn:ComptonVacuum}) has acquired additional terms in the external momenta, so its analytic structure differs from the terms above (as we will confirm more explicitly below). It may be checked that $\mcM_\mcX$ is gauge invariant.

The fourth term in (\ref{eqn:AmplitudeDecomposition}) is similar, picking up poles at  $v= v_\star$ via the $\delta$-function in the annihilation vertex:
\begin{equation}\label{eqn:VacuumPairs}
\mcM_{\mcY}
=
\frac{ 
	i
}{P_{12}^2}
\int
\!
\ud y^\LCp
\;
e^{
	i
	\Phi_\mcY(y^\LCp; v_\star)
}
\bigg[
\frac{P_{34}^2}{P_{34}^2 - P_{12}^2}
\Delta_\mcY(y^\LCp)
\mcY_0
+ 
\mcY(y^\LCp)
\bigg]
\cdot
\mcX_0
\;.
\end{equation}
Here the pair production vertex lies inside the field, while free-space momentum conservation at free annihilation vertex determines the internal photon momentum to be $\ell = P_{12}$. As such the dependence on external momenta differs to that of the previous terms.

The fifth and sixth terms $\mcM^\uparrow$ and $\mcM^\downarrow$ in (\ref{eqn:AmplitudeDecomposition}) also pick up contributions from $v = v_\star$ and $v= \bar{v}_\star$, respectively, though this time from the poles in the gauge invariant sub-amplitudes, i.e.~from within the square brackets of (\ref{eqn:RegularisedAmplitude}). These terms are, now dropping the ``$+$'' superscripts on lightfront time when unambiguous, 
\begin{align}
\label{eqn:ComptonPairsCausal}
\mcM^\uparrow &=
-
\frac{1}{2 n \cdot \ell_\star}
\int
\!
\ud y
\, e^{
	i
	\Phi_\mcY(y;v_\star)
}
\bigg[
\frac{ P_{34}^2 }{P_{34}^2  - P_{12}^2}
\Delta_\mcY(y)
\mcY_0
+ 
\mcY(y)
\bigg]
\cdot
\mcX_0 
\int^{y}
\!
\!
\!
\ud x
\,
\Delta_\mcX(x)
e^{
	i
	\Phi_\mcX(x;v_\star)
}
\;,
\\
\label{eqn:ComptonPairsAcausal}
\mcM^\downarrow &=
\frac{1}{2 n \cdot \ell_\star}
\int
\!
\ud x \,
e^{i \Phi_\mcX(x; \bar{v}_\star)} 
\bigg[
\frac{P_{12}^2}{P_{12}^2 - P_{34}^2}
\Delta_\mcX(x)\mcX_0
+ \mcX(x)\bigg]
\cdot\mcY_0
\int^{x}
\!
\!
\!
\ud y
\;
\Delta_\mcY(y)
e^{
	i
	\Phi_\mcY(y; \bar{v}_\star)
}
\;,
\end{align}
The internal line is off-shell in both cases. Both terms are (lightfront) time-ordered. In (\ref{eqn:ComptonPairsCausal}) annihilation occurs causally before pair production, while in  (\ref{eqn:ComptonPairsAcausal}) pair production occurs before annihilation\footnote{The appearance of this term in combination with lightfront time-ordering is unusual; it is an example of a ``vacuum'' diagram where the total outgoing $n\cdot p$ momentum at the pair production vertex is zero, which in lightfront quantisation, using lightfront gauge, is expected to vanish~\cite{Brodsky:1997de,Heinzl:2000ht}. This term is though gauge invariant; we will show how to recover lightfront results later.}. Observe that in both (\ref{eqn:ComptonPairsCausal}) and  (\ref{eqn:ComptonPairsAcausal}) the integrands vanish outside the of the sandwich wave,  so each interaction must occur within the field, but unlike $\mcM^\text{on}$ the vertices are not symmetric in their structure. Consider $\mcM^\uparrow$, in which annihilation occurs first.  The internal photon has momentum $\ell = P_{12}$, as it did in $\mcM_\mcX$ where the annihilation vertex was free. Here the annihilation vertex is not free, but nor is it fully dressed by the background, instead we have only
\begin{equation}\label{eqn:DressedComptonPartial}
\mcX_0^\mu 
\Delta_\mcX(x)
e^{
	i
	\Phi_\mcX(x;v_\star)
}
\;,
\end{equation}
in which the spin/polarisation structure is free, but the phase and scalar factor $\Delta_\mcX$ see the background. Despite this, both $\mcM^\uparrow$ and $\mcM^\downarrow$ are individually gauge invariant. This prompts the question of exactly what kind of interaction this vertex describes. We will give the answer in Sect.~\ref{SECT:DISCUSS}, but first we wish to make more clear the connection between the virtuality poles and the analytic structure of the amplitude as a function of \textit{external} momenta. This is most easily done by taking the perturbative limit.

\subsection{LO perturbative expansion:  poles in external momenta}\label{sec:QEDpert}
\begin{figure}
	\centering
	\includegraphics[width=0.45\textwidth,trim={3.5cm 19cm 8.8cm 4cm},clip=true]{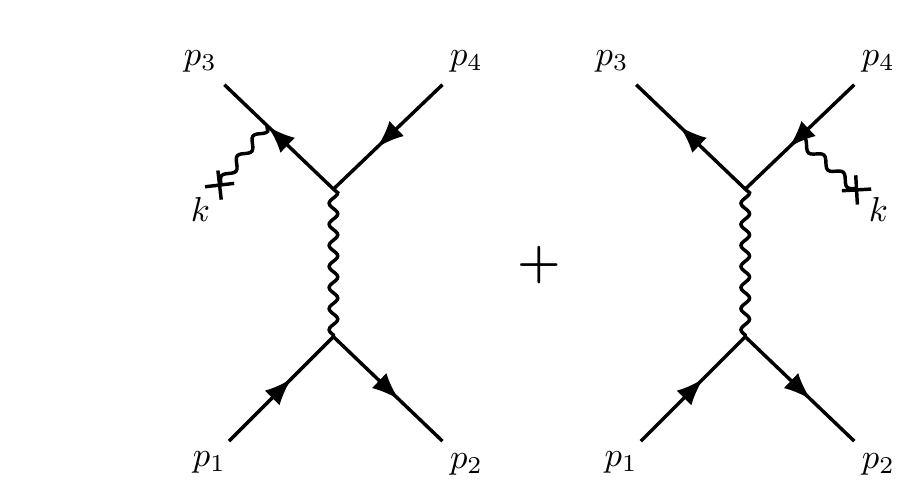}\qquad\qquad\qquad
	\caption{\label{fig:PerturbativeQED} Leading order perturbative contribution to $\mcM_\mcY$ (\ref{eqn:VacuumPairs}). Our decomposition groups together the two five-point diagrams required to maintain gauge invariance.
	}
\end{figure}

Here we show explicitly that the decomposition
(\ref{eqn:AmplitudeDecomposition}) given by the internal momentum poles splits
the amplitude into parts with different poles in the external momenta. To do so
we expand to leading order (LO) in the background. It is easily verified that
the LO contributions to $\mcM$ are linear in $a_\mu$ and come from those terms
with one background-free vertex, $\mcM_\mcX$ in (\ref{eqn:ComptonVacuum}) and
$\mcM_{\mcY}$ in (\ref{eqn:VacuumPairs}). These must correspond to some
five-point perturbative amplitude as on the right of Fig.~\ref{fig:Trident}. Expanding e.g.~(\ref{eqn:VacuumPairs}), the LO contribution is easily extracted and most conveniently written in terms of the Fourier transform $\tilde{a}_\mu$ of the field with respect to $x^\LCp$. Defining also the Fourier frequency $\omega_\star := \bar{v}_\star - v_\star$ and $k_\mu = \omega_\star n_\mu$, the LO contribution to $\mcM_\mcY$, call it $\mcM_{\mcY(1)}$, is
\begin{equation}
\label{eqn:ComptonVacuumPert}
\mcM_{\mcY(1)}
=
i
\bar{u}_{p_3}
\bigg[
\frac{
	\slashed{\tilde{a}}(\omega_\star)
	\big(
	\slashed{p_3}
	-
	\slashed{k}
	+
	m
	\big)
	\gamma^\mu
}{(p_3 - k)^2 - m^2}
+
\frac{
	\gamma^\mu
	\big(
	\slashed{k}
	-
	\slashed{p_4}
	+
	m
	\big)
	\slashed{\tilde{a}}(\omega_\star)
}{(p_4 - k)^2 - m^2}
\bigg]
v_{p_4}
\,
\frac{1}{(p_1 + p_2)^2}
\,
\bar{v}_{p_2}
\gamma_\mu
u_{p_1}
\;.
\end{equation}
The pair annihilation vertex is the vacuum vertex, while the pair production vertex reduces to the textbook expression for tree level pair production by two photons in vacuum, $\gamma \gamma \rightarrow e^- e^+$, with one photon convoluted with the background ${\tilde a}_\mu$. Observe that a single term in our decomposition has yielded \textit{both} interaction channels for $\gamma \gamma \rightarrow e^- e^+$, which are required for gauge invariance, see Fig.~\ref{fig:PerturbativeQED}.

An analogous calculation shows that $\mcM_{\mcX(1)}$, the LO contribution to (\ref{eqn:ComptonVacuum}), has a similar expression in which the external field couples to one of the incoming, rather than outgoing, pair. From this description it is clear that $\mcM_{\mcX(1)}$ and $\mcM_{\mcY(1)}$ must have a different analytic structure as functions of external momenta; there are poles in (\ref{eqn:ComptonVacuumPert}) at $(p_1+p_2)^2=0$, $(p_3-k)^2=m^2$ and $(p_4-k)^2=m^2$, but $\mcM_{\mcX(1)}$ has instead poles at $(p_3+p_4)^2=0$, $(p_1+k)^2=m^2$ and $(p_2+k)^2=m^2$. In the next section we will see how these structures extend to next-to-leading order (NLO). 

\section{Soft separation in background field amplitudes}\label{SECT:DISCUSS}
Compare $\mcM_\mcY$ in (\ref{eqn:VacuumPairs}) with $\mcM^\uparrow$ in (\ref{eqn:ComptonPairsCausal}). Both contain the fully dressed pair production vertex. The difference between the two is in the annihilation vertex. This is free in $\mcM_\mcY$, but in $\mcM^\uparrow$ depends on the background through the simpler vertex (\ref{eqn:DressedComptonPartial}). Comparing the two, we see we can write $\mcM^\uparrow$ as
\begin{equation}\label{eqn:exact-soft}
\mcM^\uparrow = -i \frac{P_{12}^2}{2n\cdot P_{12}} \int\!\ud y \,
{\mcM}^{\bm\prime}_\mcY	
\int^{y}
\!
\!
\!
\ud x
\,
\Delta_\mcX(x)
e^{
	i
	\Phi_\mcX(x;v_\star)
}
\;,
\end{equation}
in which ${\mcM}^{\bm\prime}_\mcY$ is shorthand for the integrand of $\mcM_\mcY$. We see that, at the level of the integrand, $\mcM^\uparrow$ is a \textit{scalar} multiple of $\mcM_\mcY$. A similar relation holds for $\mcM^\downarrow$ and $\mcM_\mcX$.  Our focus is now on the physical interpretation of this structure.

\subsection{Soft interactions with the background}
In order to understand (\ref{eqn:exact-soft}), we again turn to perturbation theory. Expanding $\Delta_\mcX$ in powers of the background, using (\ref{def:DeltaC}) and (\ref{pi-1}), we have the lowest order contribution\footnote{The neglected terms are only quadratic in $a$ and easily written down.}
\begin{equation}
\label{def:DeltaCAlt}
\Delta_\mcX(x^\LCp)
=
-
\frac{2 n \cdot P_{12}}{P_{12}^2}
a_\mu(x^\LCp)
\bigg[
\frac{p_1^\mu}{n \cdot p_1}
-
\frac{p_2^\mu}{n \cdot p_2}\bigg]
+ \ldots
\;.
\end{equation}
We recognise in the square brackets a Weinberg `soft-factor' for soft emission/absorption of background photons, characterised by direction $n_\mu$, at the pair annihilation vertex, with $a_\mu$ taking the place of the polarisation vector. The significance of this follows from observing that since both $\mcM_\mcY$ and $\mcM^\uparrow$ pick up the same pole, the internal line carries momentum $\ell = P_{12}$ in both cases; hence while there is an interaction with the background at the annihilation vertex in $\mathcal{M}^\uparrow$, this interaction does not enter the momentum conservation law. Keeping track of the different kinematic prefactors in $\mcM^\uparrow$ and $\mcM_\mcY$, the LO effect of this interaction is simply to multiply (up to Fourier transform factors) the five point amplitude $\mcM_{\mcY(1)}$ by the soft factor above, so
\begin{equation}\label{nypol1}
\mcM^\uparrow_{(2)} 
\sim
a
\cdot
\bigg[
\frac{p_1}{n \cdot p_1}
-
\frac{p_2}{n \cdot p_2}\bigg] 
\times 
\left(
\,
\raisebox{-20pt}{\includegraphics[width=0.15\textwidth,trim={5.25cm 19.7cm 10.29cm 4.42cm},clip=true]{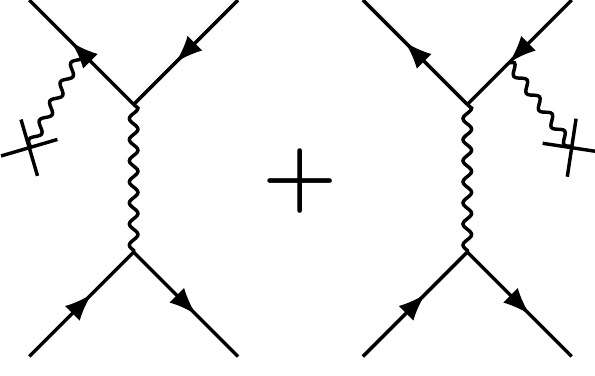}}
\,
\right)+ \ldots 
\end{equation}
This is explicitly a hard-soft factorisation; the hard part of the process is the perturbative five-point amplitude (\ref{eqn:ComptonVacuumPert}), Fig.~\ref{fig:PerturbativeQED}, in which the external field couples as normal to the created pair, while the soft factor describes emission/absorption of background photons at the annihilation vertex. The soft factor also affects the analytic structure; relative to $\mcM_{\mcY(1)}$, there are in $\mcM^{\uparrow}_{(2)}$ additional poles at $n\cdot p_1=0$, $n\cdot p_2=0$.  Analogous results hold for $\mcM^\downarrow_{(2)}$ and $\mcM_\mcX$ which both pick up the pole at $v=\bar v_\star$ such that the internal momentum is $\ell = p_3 + p_4$. The hard-soft factorisation is
\begin{equation}
\mcM^\downarrow_{(2)} 
\sim 
a\cdot
\bigg[
\frac{p_3}{n \cdot p_3}
-
\frac{p_4}{n \cdot p_4}\bigg] \times
\left(
\,
\raisebox{-20pt}{\includegraphics[width=0.15\textwidth,trim={5.25cm 19.7cm 10.29cm 4.42cm},clip=true]{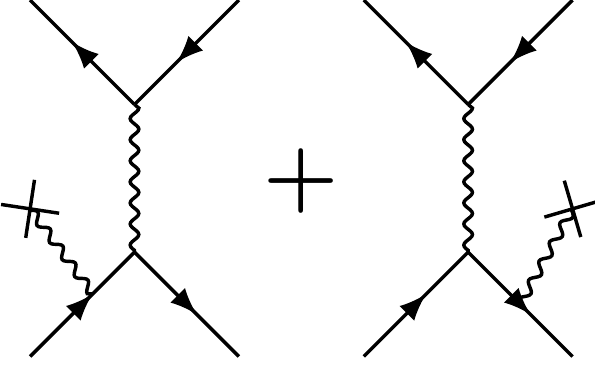}}
\,
\right)
+
\ldots 
\end{equation}
with the poles in $\mcM^\uparrow$ obtained from $\mcM^\downarrow$ by exchanging $\{p_1,p_2\}$ for $\{p_3,p_4\}$.

Beyond these lowest order calculations, it remains true that the momentum is unchanged at the vertices of the type (\ref{eqn:DressedComptonPartial}). Thus their only effect is to introduce (under the lightfront time integral) a scalar factor which, perturabtively, is a standard soft emission factor. The interpretation of (\ref{eqn:exact-soft}) is then that it gives an all-orders hard/soft factorisation in our background, which holds locally (i.e~under the integral) because of the nontrivial spacetime dependence introduced by the background. It would be interesting to connect this to inverse-soft theorems~\cite{ArkaniHamed:2009dn,Nguyen:2009jk,BoucherVeronneau:2011nm,Nandan:2012rk}.

\begin{figure}[t!!]
	\centering
	\begin{subfigure}[b]{0.3\textwidth}
		\centering
		\includegraphics[width=1.1\textwidth,trim={3.8cm 19cm 11cm 4cm},clip=true]{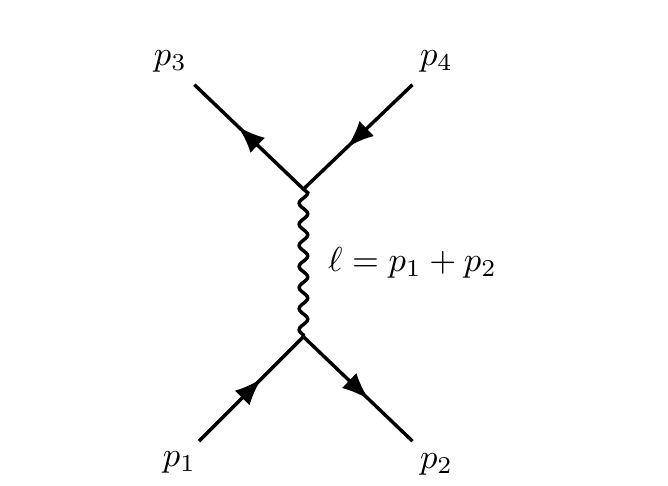}
		\caption{$\mcM^{\text{vac}}$}
		\label{fig:f}
	\end{subfigure}
	\begin{subfigure}[b]{0.3\textwidth}
		\centering
		\includegraphics[width=1.1\textwidth,trim={3.8cm 19cm 11cm 4cm},clip=true]{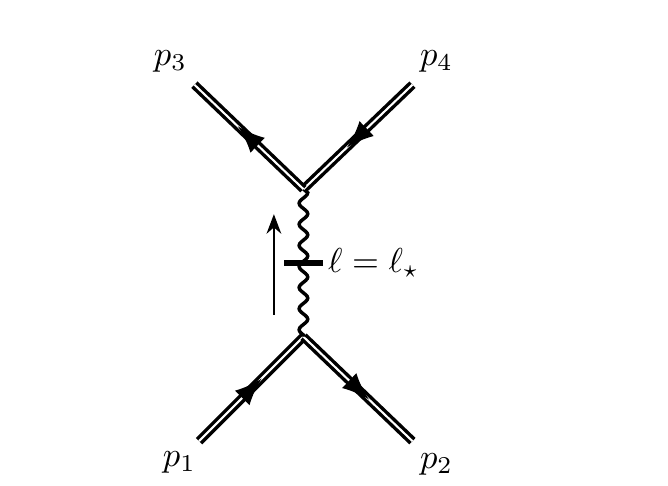}
		\caption{$\mcM^{\text{on}}$}
		\label{fig:e}
	\end{subfigure}
	\begin{subfigure}[b]{0.3\textwidth}
		\centering
		\includegraphics[width=1.1\textwidth,trim={3.8cm 19cm 11cm 4cm},clip=true]{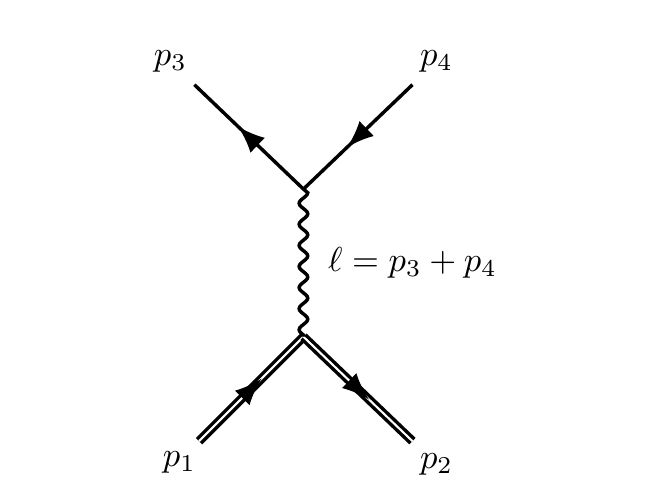}
		\caption{$\mcM_{\mcX}$}
		\label{fig:a}
	\end{subfigure}
	\begin{subfigure}[b]{0.3\textwidth}
		\centering
		\includegraphics[width=1.1\textwidth,trim={3.8cm 19cm 11cm 4cm},clip=true]{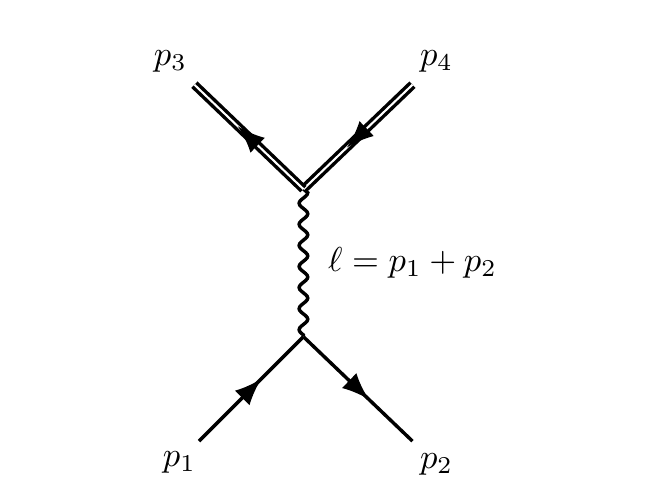}
		\caption{$\mcM_{\mcY}$}
		\label{fig:b}
	\end{subfigure}
	\begin{subfigure}[b]{0.3\textwidth}
		\centering
		\includegraphics[width=1.1\textwidth,trim={3.8cm 19cm 11cm 4cm},clip=true]{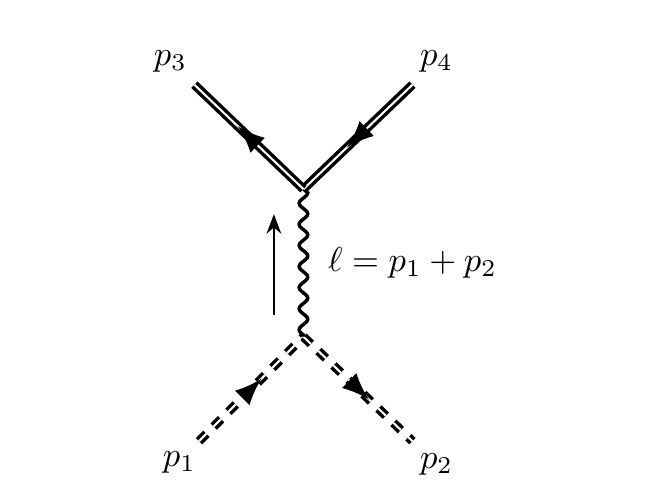}
		\caption{$\mcM^{\uparrow}$}
		\label{fig:d}
	\end{subfigure}
	\begin{subfigure}[b]{0.3\textwidth}
		\centering
		\includegraphics[width=1.1\textwidth,trim={3.8cm 19cm 11cm 4cm},clip=true]{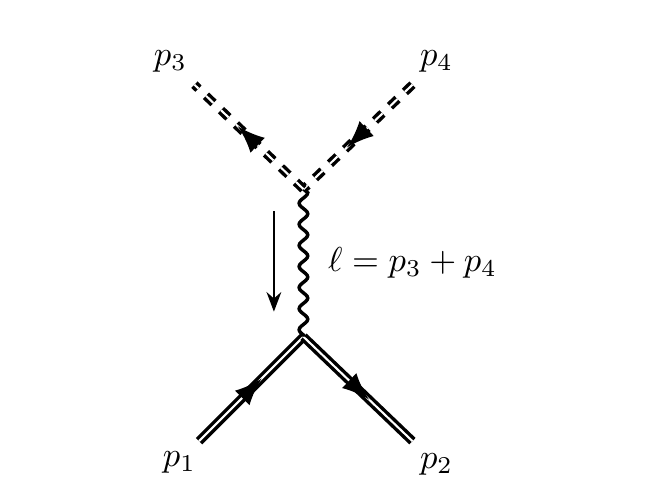}
		\caption{$\mcM^\downarrow$}
		\label{fig:c}
	\end{subfigure}
	\caption{The decomposition (\ref{eqn:AmplitudeDecomposition}) of the scattering amplitude $\mcM$ into gauge invariant pieces. Arrows denotes the momentum flow through the propagator. Dashed lines indicate the soft dressing. The cut in Fig. \ref{fig:e} indicates that the intermediate photon is on-shell, $\ell = \ell_\star$ with $\ell_\star^2 = 0$.
	}
	\label{fig:Decomposition}
\end{figure}

In conclusion, our decomposition of the full scattering amplitude, into terms with different internal poles, also corresponds to a separation into hard and soft parts in terms of the external momenta. These results hint at an underlying structure and classification of how a background can interact with particles, or ``dress'' a vertex. We have seen three types of interaction:
\begin{enumerate}
	\item \textit{No interaction with the background}: the vertex is exactly equal to the vacuum expression, with no influence of the background on the fermions at that vertex.  The intermediate photon is off-shell, with the virtuality determined by (background-free) conservation of \textit{four}-momentum.
	
	\item \textit{Soft interaction}: the background affects the interaction at a vertex, but only `softly': the only contribution is a soft factor. There is in particular no contribution to the momentum flow at the vertex.  We  refer to such vertices as soft.
	
	\item \textit{Hard interaction}: the fully dressed vertex appears, the interaction with the background affects the momentum flow through the vertex, and the tensor structure is not simply a soft factor, and only \textit{three}-momentum is conserved.
\end{enumerate}
In terms of the these three, a diagrammatic representation of each of the sub-amplitudes in (\ref{eqn:AmplitudeDecomposition}) is shown in Fig. \ref{fig:Decomposition}. Interactions at hard (fully dressed) vertices are indicated by solid double lines as above, vacuum vertices by single lines, and soft interactions by dashed double fermion lines. Each of these diagrams is individually gauge invariant. The only term with two `hard' vertices is the on-shell term, implying absorption of energy from the background at both vertices. Physically this makes sense; each term in the amplitude factorises at a different virtuality, and for the on-shell pole, neither of the three-point sub-amplitudes can occur in vacuum with all particles on-shell unless assisted by the background. 
%

%
%
%
\section{Scalar Yukawa and the infra-red \label{sec:Scalar}}
%
We have seen that gauge invariance of QED amplitudes is intimately related to their infra-red, or soft, behaviour. Soft limits can determine the analytic structure of amplitudes in theories without gauge symmetry~\cite{Rodina:2018pcb}. We therefore consider here a simple scalar Yukawa theory, and show that analogous analytic structures to those in QED emerge from the soft behaviour of amplitudes. 
We consider a scalar Yukawa theory of a massive `electron' $\varphi$, massless `photon'~$A$, and external field $A_\text{ext}$, 
\begin{equation}\label{SKL}
\mathcal{L} = \frac{1}{2}\big(\partial\varphi \cdot \partial\varphi - m^2 \phi^2\big) + \frac{1}{2} \partial A\cdot \partial A - g \varphi^2 (A + A_\text{ext})\;,
\end{equation}
in which the coupling $g$ has mass dimension one in four dimensions. Since the Feynman rules of the theory mimic those of QED we will here be able to reinforce the preceding results in a technically simpler setting. The external sandwich wave is now $g A_\text{ext}(x) = a(x^\LCp)$, which has mass dimension $2$. In analogy to QED, incoming electron legs are represented by
\begin{equation}\label{skalar-ben}
\varphi_p(x)  = \exp\bigg[-ip\cdot x - \frac{i}{2n.p}\int\limits_{-\infty}^{x^\LCp} \! \ud s\, a(s)\bigg]  \;,
\end{equation}
where $p^2=m^2$. For outgoing electrons $\varphi_p^\dagger$ take the conjugate and replace $-\infty \to +\infty$ in the exponent. In analogy to QED, a kinetic momentum $\pi_\mu$ can be defined as
\begin{equation}\label{def:pi-scalar}
\pi_\mu(x^\LCp) = p_\mu + \frac{a(x^\LCp)}{2n.p} n_\mu \;,
\end{equation}
which obeys $\pi^2(x^\LCp) = m^2 +a(x^\LCp)$; this is the classical mass-shell condition, because in (\ref{SKL}) the background is equivalent to a spacetime-dependent mass.

\subsection{Infra-red behaviour}

We again focus on the $2\to 2$ `electron' scattering amplitude in Fig.~\ref{fig:Trident}. Writing $iG$ for the scalar photon propagator, the $S$-matrix element is
\begin{equation}
\label{SFI-scalar}
\begin{split}
S_\text{fi} 
&= 
-i g^2
\int\!\ud^4 y
\!
\int\!\ud^4x
\, 
\varphi_{p_3}^\dagger(y) \varphi_{p_4}^\dagger(y) G(y-x) \varphi_{p_2}(x)\varphi_{p_1}(x) \quad +\,\,\, \cdots \\
&= -ig^2 (2\pi)^3\delta^3_\text{LF}(p_1 + p_2 - p_3 - p_4)  \mathcal{M} \quad + \,\,\, \cdots  \;,
\end{split}
\end{equation}
in which the ellipses denote permutations of external legs etc and $\mathcal{M}$ is the reduced amplitude obtained by integrating out the transverse and longitudinal coordinates. The intermediate photon momentum is again $\ell^\mu =\ell_\star^\mu + v n^\mu$ with $\ell_\star$ as defined in (\ref{def:lstar}), and $\mathcal{M}$ may be written as an integral over the virtuality $v$,
\begin{equation}
\mathcal{M} = \frac{i}{2n\cdot \ell}\int\!\frac{\ud v}{2\pi}\frac{1}{v+i\epsilon}
\int \! \ud y^\LCp
e^{i \Phi_{\mcY}(y^{\LCp};v)}
\int \! \ud x^\LCp
\,  
e^{i \Phi_{\mcX}(x^{\LCp};v)}
\;.
\end{equation}
The functions in the exponents, $\Phi_{\mcX}(x^{\LCp};v)$ and $\Phi_{\mcY}(y^{\LCp};v)$ are given by (\ref{def:ComptonPhase}) but with the kinetic momenta given by $\hat\pi\to\pi\to$ (\ref{def:pi-scalar}). The integrand at each vertex integral in $\mathcal{M}$ is a pure phase, the IR behaviour of which is not explicit. An entirely analogous calculation to that in QED, in which we introduce damping factors and identify the IR contributions, leads to the regularised expression, once again dropping $+$ subscripts on lightfront times,
\begin{align}
&\mathcal{M} \to 	\frac{i}{2n\cdot \ell}\int\!\frac{\ud v}{2\pi} \frac{1}{v+i\epsilon} \\
&\nonumber\bigg[ 2\pi \delta(v-\bar v_\star) - \frac{\bar v_\star}{v-\bar v_\star-i\epsilon}\int\! \ud y \, \mcY(y,v)\bigg] 
\bigg[  2\pi \delta(v-v_\star) - \frac{v_\star}{v-v_\star+i\epsilon} \int\!\ud x\, \mcX(x,v)\bigg] \;.
\end{align}
in which there are new poles in $v_\star$ and $\bar v_\star$ with the same definitions as in QED, (\ref{def:DeltaC}) and (\ref{def:tstar}). The structure of the amplitude is very similar to that of QED, reflecting the universality of soft behaviour. The vertex functions $\mcX$ and $\mcY$ may be conveniently written as
\begin{align}\label{skal-full}
\mcY(y,v) &= \Delta_\mcY(y) e^{i \Phi_{\mcY}(y;v)} \;, 
\quad&
\mcX(x,v) &= \Delta_\mcX(x) e^{i \Phi_{\mcX}(x;v)} \;,
\end{align}
where the $\Delta$ factors have the same form as (\ref{def:DeltaC}) and (\ref{def:tstar}) but with $\hat\pi\to\pi\to$ (\ref{def:pi-scalar}).

Performing the virtuality integral and picking up the pole contributions we obtain six terms which correspond exactly to the QED decomposition (\ref{eqn:AmplitudeDecomposition}). The term $\mcM^\text{vac}$ from the product of delta-functions is nothing but the background-free contribution, yielding
\begin{equation*}
S_{fi} = ig^2 (2\pi)^4 \delta^4\big(P_{12} - P_{34}\big)\frac{1}{P_{12}^2} \;.
\end{equation*}
The on-shell term depends on the on-shell momentum $\ell_\star$ and is time-ordered as before,
\begin{equation}\label{skal-on}
\mcM^\text{on} = \frac{1}{2n\cdot \ell_\star} \int\!\ud y \!\!\int^{y}\!\!\!\ud x\,  \mcY(y,0) \mcX(x,0) \;.
\end{equation}
The analogues of $\mcM_\mcY$ and $\mcM_\mcX$ in which one vertex lies outside the field are
\begin{equation}
\mcM_\mcY =  \frac{i P_{34}^2}{P_{12}^2 (P_{34}^2-P_{12}^2)} \int\!\ud y\, \mcY(y,v_\star)  \;, \quad 
\mcM_\mcX = \frac{-i P_{12}^2}{P_{34}^2(P_{34}^2-P_{12}^2)} \int\!\ud x\, \mcX(x,\bar{v}_\star)
\;.
\end{equation}
The vacuum vertices are simply factors of unity here, which obscures their identification compared to QED. However, we can see in the argument of the photon absorption vertex $\mcY$ that the intermediate photon carries the momentum $\ell = P_{12}$ which would be assigned by the vacuum annihilation vertex (and vice versa for $\mcX$). The remaining terms in our expansion are
\begin{align}
\mcM^\uparrow&= - \frac{1}{2n\cdot \ell_\star} \frac{P_{34}^2}{P_{34}^2-P_{12}^2} \int\!\ud y\mcY(y,v_\star)\! \int^{y}\!\!\!\ud x\,\mcX(x,v_\star)  \label{skal-kausal}\;, \\
\mcM^\downarrow &= - \frac{1}{2n\cdot \ell_\star} \frac{P_{12}^2}{P_{34}^2 - P_{12}^2} \int\!\ud x \mcX(x,\bar{v}_\star) \!\!\int^{x}\!\!\!\ud y\,  \mcY(y,\bar{v}_\star) \;. \label{sista}
\end{align}
The same time ordering is present as in QED, with the pair annihilation vertex occurring first (second) in $\mcM^\uparrow$ ($\mcM^\downarrow$). Note that the analogue of the QED `soft' vertex is, here, the full vertex (\ref{skal-full}), because we have no spin of polarisation, which makes the hard-soft factorisation we saw in QED less explicit; it remains nevertheless, as the momentum assigned to the internal line in $\mcM^\downarrow$ and $\mcM^\uparrow$ is the same background-free assignment as in $\mcM_\mcY$  and $\mcM_\mcX$ respectively, and the scalar-multiple relation (\ref{eqn:exact-soft}) is clear in (\ref{skal-kausal})--(\ref{sista}).

As for QED, the additional poles in the \textit{internal} momentum have factorised our amplitude into parts with different analytic structure in the \textit{external} momenta -- this will be made explicit by examining the perturbative structure of the amplitudes in the following two subsections.  We first note that the $\Delta$ factors in this scalar setting have a simpler form; they are almost scalar soft factors multiplied by $a$:
\begin{align}
\Delta_\mcY(y) &=  -\frac{a(y) }{\bar v_\star} \bigg(\frac{1}{2n\cdot p_3}+ \frac{1}{2n\cdot p_4}\bigg) =: - \frac{a(y)}{\bar v_\star}  W_{34} \\
\Delta_\mcX(x) &= \frac{a(x) }{v_\star} \bigg(\frac{-1}{2n\cdot p_2}+ \frac{-1}{2n\cdot p_1}\bigg) =: \frac{a(x)}{v_\star}  W_{12} \;.
\end{align}
In a moment we will see how the missing momentum scale in $W_{34}$ and $ W_{12}$ is assigned, changing them into soft factors proper.

\subsection{Comparison with LO perturbation theory}
The lowest order perturbative contribution is again $\mcO(a_0)$, and comes from $\mcM_\mcY$ and $\mcM_\mcX$ in which one vertex is background-free. To this order, we may set $a\to 0$ in the exponentials. The lightfront time integral then gives the Fourier transform of $a$ appearing in the $\Delta$ factor. The reduced amplitude becomes, writing $\omega_\star \equiv \bar v_\star - v_\star$, 
\begin{equation}\label{skal-a0-1-1}
\mcM \to \mcM_{\mcY(1)} + \mcM_{\mcX(1)}  =  -i \frac{2n\cdot \ell_\star}{P_{34}^2-P_{12}^2} \,
\tilde{a}\big(\omega_\star\big)\, \bigg[\frac{ W_{34}}{P_{12}^2}  +  \frac{ W_{12}}{P_{34}^2} \bigg] \;.
\end{equation}
The first term in (\ref{skal-a0-1-1}) comes from $\mcM_\mcY$ and corresponds to the \textit{pair} of diagrams in Fig.~\ref{fig:PerturbativeQED}. The second term in (\ref{skal-a0-1-1}) comes from $\mcM_\mcX$ and corresponds to the pair of diagrams with the external field photon attached to incoming legs. Noteably, IR behaviour \textit{groups} emission from the outgoing electrons, and emission from the incoming electrons, together, just as happens in QED where it is necessary for gauge invariance.

We now write $\tilde{a}$ as (trivially) an integral over frequencies $\ud \omega$ weighted with a delta function fixing $\omega \to \omega_\star$. This delta-function combines with that in the prefactor to recover the covariant delta-function of a perturbative \textit{five}-point amplitude describing the scattering of the original set of matter particles and an additional photon of momentum $k_\mu \equiv \omega n_\mu$. This momentum defines the soft factors $\tilde W $ proper,
\begin{equation}
{\tilde W}_{34} = \frac{1}{2k\cdot p_3}+ \frac{1}{2k\cdot p_4}  \;, \qquad {\tilde W}_{12} = \frac{-1}{2k\cdot p_2}+ \frac{-1}{2k\cdot p_1} \;,
\end{equation}
and allows us to simplify (\ref{skal-a0-1-1}); the corresponding $S$-matrix element is
\begin{equation}\label{skal-a0-1-2}
\mathcal{S}_\text{fi} =i g^2 \int\!\frac{\ud \omega}{2\pi}\,  \tilde{a}(\omega)\,\, (2\pi)^4\delta^4(P_{34} - P_{12} - k)\, \, \bigg[\frac{{\tilde W}_{34}}{P_{12}^2}  + \frac{{\tilde W}_{12}}{P_{34}^2} \bigg] + \cdots \;.
\end{equation}
This is precisely the tree level contribution to the scalar five-point amplitude $e+e+k \to e + e$, with the photon momentum convoluted with the field profile. 

\subsection{Expansion to NLO \label{sec:ScalarNLO}}
At $\mcO(a_0^2)$ our expressions depend on the soft factors $W$ and on a Fourier transform factor $F$, which is now quadratic in the field, defined by
\begin{equation}
F(\alpha , \beta ) := \int\!\ud y \!\int\!\ud x \, \theta(y-x) \, e^{i\alpha y} a(y) \, e^{-i \beta x} a(x) \;.
\end{equation}
The on-shell term becomes (a subscript $(2)$ denotes second order in perturbation theory)
\begin{equation}
\mcM^\text{on}_{(2)} = \frac{-2n\cdot \ell_\star}{P_{34}^2 P_{12}^2} W_{34} W_{12}  F(\bar v_\star,v_\star) \;.
\end{equation}
in which the soft factors $W$ come directly from the $\Delta$ factors. For the terms with one vertex outside the field, the soft factors at second order come both from $\Delta$ and from expanding the phases; we find
\begin{equation}
\mcM_{\mcX (2)} = \frac{2n\cdot \ell_\star}{P_{34}^2(P_{34}^2-P_{12}^2)}W_{12}^2 F(\omega_\star,0)\;, \qquad \mcM_{\mcY (2)} = \frac{2n\cdot \ell_\star}{P_{12}^2(P_{34}^2-P_{12}^2)}W_{34}^2 \, F(\omega_\star,0) \;.
\end{equation}
Note both the different denominators and soft factors compared to the on-shell term. The different Fourier factor reflects the fact that no energy-momentum is taken from the background at one of the vertices. Finally, the scalar analogue the sub-amplitudes with one hard and one soft vertex are 
\begin{equation}
\mcM^\uparrow_{(2)} = \frac{2n\cdot \ell_\star}{P_{12}^2(P_{34}^2-P_{12}^2)}W_{34}W_{12} \, F(\omega_\star,0) \;, \quad 
\mcM^\downarrow_{(2)} =  \frac{2n\cdot \ell_\star}{P_{34}^2(P_{34}^2-P_{12}^2)}W_{34}W_{12} \, F(\omega_\star,0) \;.
\end{equation}
From this we can exhibit the analogue of the hard/soft factorisation found in QED. The second order contributions $\mcM_{(2)}^\uparrow$ and $\mcM_{(2)}^\downarrow$ are \textit{six-point} amplitudes in perturbation theory. They are given, up to Fourier transform factors, by multiplying the \textit{five-point} amplitudes $\mcM_{\mcY(1)}$ and $\mcM_{\mcX(1)}$ by soft factors $W_{12}$  and $W_{34}$ respectively: 
\begin{equation}
\mcM^\uparrow_{(2)} = i \frac{F(\omega_\star,0)}{\tilde{a}\big(\omega_\star\big)} W_{12}  \mcM_{\mcY(1)}  \;,
\qquad		
\mcM^\downarrow_{(2)}=   i \frac{F(\omega_\star,0)}{\tilde{a}\big(\omega_\star\big)} W_{34}  \mcM_{\mcX(1)} \;.
\end{equation}
Each of these terms has, accounting for the soft factors, a different functional dependence on, and different poles in, the external momenta. The terms are grouped in the same way as the gauge invariant  QED groupings. All terms in which the photon is off-shell share the same $F$ factor, which differs from the on-shell term.

%
%
%
\section{Conclusions \label{sec:outro}}
%
It has been shown for several  theories that gauge invariance and soft limits are enough to determine the analytic structure of scattering amplitudes. We have made a connection between these results and QED scattering on background plane waves, showing that imposing explicit gauge invariance reveals a previously obscured analytic structure in scattering amplitudes. Gauge invariance introduces new poles into the virtuality integral of internal lines. Amplitudes factorise at each of these poles, giving a new decomposition in which each term is individually gauge invariant and has a different analytic structure in the \textit{external} scattering momenta.

Further, we saw that gauge invariance was closely linked to the infra-red behaviour of amplitudes, and that the resulting decomposition separated out terms with a \textit{soft} interaction with the background, resulting in a decomposition into background-free, soft, and hard interactions with the background. This connection with the infra-red allowed us to extend our results to a simple scalar Yukawa theory. Exposing the infra-red behaviour of the scalar amplitudes resulted in a very similar decomposition to that in QED, with each term in the decomposition having a different analytic structure.

We remark that the decomposition of amplitudes into gauge invariant
sub-amplitudes, both here and more generally, is reminiscent of two different
approaches; the ``\emph{pinch technique}'' in QCD~\cite{Cornwall:1976hg} and the
``\emph{background field method}''~\cite{Abbott:1981ke}.  In the pinch technique
a cancellation of gauge dependent terms~\cite{Lavelle:1991ve} when going from
correlation functions to scattering amplitudes occurs in such a way as to
decompose amplitudes into kinematically distinct, individually gauge-invariant
sub-amplitudes.  See~\cite{Binosi:2009qm} for a review.  The background field
approach is used to derive effective actions in a manifestly gauge invariant way
by perturbing a quantum field around a classical background. It has been used as
an alternative to the pinch technique, with both agreeing to one
loop~\cite{denner1994}.  It would be interesting to investigate how these
approaches are related to the work presented here, along with possible
connections between the structures in our amplitudes and inverse-soft
theorems~\cite{ArkaniHamed:2009dn,Nguyen:2009jk,BoucherVeronneau:2011nm,Nandan:2012rk}.
We leave this to future work.

A natural question for future work is whether gauge invariance can be applied constructively to fully determine amplitudes in background fields. We also wish to establish more firmly the universality of our results. At the level of four point functions (which is often enough to reveal new structure~\cite{Carrasco:2019yyn}), we should also consider processes with an intermediate fermion dressed by the background. Rather than pursue this in QED, we will instead consider Yang Mills and QCD in plane waves, following~\cite{Adamo:2017nia,Adamo:2018mpq,Adamo:2019zmk}, in which case all particles, both massless and massive, are dressed. Higher $N$-point amplitudes will also be investigated. We hope our results will help in understanding the on-shell construction of the electroweak sector of the standard model~\cite{Durieux:2019eor,Bachu:2019ehv};  we have seen hints that the deep connections between gauge invariance, the infra-red, and analytic structure of scattering amplitudes may be found in \textit{general} theories. \\

\textit{The authors thank Tim Adamo for useful discussions and comments on a draft of this paper. The authors are supported by EPSRC, grant EP/S010319/1.}  %
%
\appendix
\section{Trident pair production\label{app:Trident}}
%
The large distance regularisation used above is standard when discussing infra-red effects~\cite{Peskin:1995ev,Weinberg:1995mt} and is well-known in the literature on QED in strong plane wave backgrounds (in which `strong' refers to the regime $a/m > 1$ whereupon the coupling to the background cannot be treated perturbatively).  In the context of three-point amplitudes it was used as a method to remove seemingly unphysical contributions to the amplitude from the spacetime region outside the sandwich background~\cite{Boca09}. However, our results show that this interpretation does not hold higher $N$-point amplitudes; in the decomposition (\ref{eqn:AmplitudeDecomposition}) there are terms $\mcM_\mcX$ and $\mcM_\mcY$ in which one vertex can lie outside the background. That the procedure removes such contributions from three-point amplitudes is thus largely coincidental; as we have seen, what the regularisation is really doing is imposing gauge invariance.

It has even been recognised, for three-point~\cite{Dinu:2012tj} and four-point amplitudes~\cite{ilderton11} that gauge invariance \textit{implies} the relation between parts of sub-amplitudes which follows from the infra-red regularisation. However, for three-point amplitudes there is no free virtuality parameter $v$, so it was not recognised that the regularisation would introduce poles into higher point amplitudes. For four-point amplitudes, most authors perform the virtuality integral before considering gauge invariance~\cite{Hu:2010ye,ilderton11,dinu18,mackenroth18}, hence the existence of the additional poles, and the structure they reveal, was not previously noticed.  (The closest to our approach is in~\cite{acosta19}, where similar expressions for the reduced amplitudes in trident appear, however the effect of the regularisation on the analytic structure of the amplitude was not recognised.) 

This prompts us to make a more explicit connection to the existing literature. By making the change $\bar{\Psi}_{-p_2} \rightarrow \bar{\Psi}_{p_2}$ in (\ref{eqn:TridentAmplitude}) we obtain the amplitude for trident pair production, $e^\LCm \to e^\LCm + e^\LCm + e^\LCp$.  We saw above that $\mcM_\mcX$ and $\mcM^\downarrow$ pick up contributions at the same virtuality (as do $\mcM_\mcY$ and $\mcM^\uparrow$); if we add these terms together, an integration by parts shows that our expressions for trident match those in~\cite{mackenroth18}, though in doing so we lose the hard-soft factorisation, and separation into different analytic structures. The results of \cite{mackenroth18} were checked to be equal to those in~\cite{dinu18} calculated previously in a different gauge. Thus, our approach reproduces literature representations of the trident process.

\bibliographystyle{JHEP}

\providecommand{\href}[2]{#2}\begingroup\raggedright\endgroup

\end{document}